\documentstyle[11pt]{article}
\begin{document}
\title{\Large \bf Soliton Solutions on Noncommutative Orbifold $ T^2/Z_4 $}
\author{ Hui Deng $$ \thanks{Email:hdeng@phy.nwu.edu.cn},
\hspace{5mm}
         Bo-Yu Hou $$ \thanks{Email:byhou@phy.nwu.edu.cn},
\hspace{5mm}
        Kang-Jie Shi$$ \thanks{Email:kjshi@phy.nwu.edu.cn},
                         \hspace{5mm}
        Zhan-Ying Yang $$ \thanks{Email:yzy@phy.nwu.edu.cn},
                          \hspace{5mm}
           Rui-Hong Yue$$  \thanks{Email:yue@phy.nwu.edu.cn}\\[1cm]
        Institute of Modern Physics, Northwest University,\\
        Xi'an, 710069, P. R. China}
\maketitle
\vspace{1cm}
\begin{abstract}
In this paper, we explicitly construct a series of projectors on
integral noncommutative orbifold $T^2/Z_4$ by extended $GHS$
constrution. They include integration of two arbitary functions
with $Z_4$ symmetry. Our expression possess manifest $Z_{4}$
symmetry. It is proved that the expression include all projectors
with minimal trace and in their standard expansions, the eigen
value functions of coefficient operators are continuous with
respect to the arguments $k$ and $q$. Based on the integral
expression, we alternately show the derivative expression in terms
of the similar kernal to the integral one. Since projectors
correspond to soliton solutions of the field theory on the
noncommutative orbifold, we thus present a series of corresponding
solitons.
\end{abstract}

 \medskip \medskip

\medskip

\section{Introduction}

String theory is a very promising candidate for an unified
description of the fundamental interactions, including quantum
gravity. It may provide a conceptual framework to resolve the
clash between two of the greatest achievements of 20th century
physics: general relativity and quantum mechanics. Noncommutative
geometry is originally an interesting topic in mathematics
\cite{1,2,3}, In the past few years, it has been shown that some
noncommutative gauge theories can be embedded in string theories \cite{4,5,6}%
and noncommutative geometry can also be applied to condensed
matter physics. The currents and density of a system of electrons
in a strong magnetic field may be described by a noncommutative
quantum field theory [7,8,9]. The connection between a finite
quantum Hall system and a noncommutative Chern-Simon Matrix model
first proposed by \cite{hall20} was further elaborated in papers
\cite{hall21,hall22}. Many papers are concentrated on the research
for the related questions about the quantum hall effect [12-18].
Since the noncommutative space resemble a quantum phase space, it
exhibits an interesting spacetime uncertainty relation, which
cause a $UV/IR$ mixing \cite{bib:uv1,bib:uv2} and a teleological
behavior. Noncommutative field theories can be regarded as highly
constrained deformation of local field theory. Thus it may help us
to understand non-locality at short distances in quantum gravity.

Solitons in various noncommutative theories have played a central
role in understanding the physics of noncommutative theories and
certain situations of string theories. The quantum Hall effect
practically provides a good illustration of the combination of the
three theories \cite{bib:hall2,bib:hall6,bib:hall9,bib:hall10}.
The existence and form of these classical solutions are fairly
independent of the details of the theory, making them useful to
probe the string behavior. In fact these solitons are the
(lower-dimensional) D-branes of string theory manifested in a
field theory limit while still capturing many string features.

Starting from the celebrated paper of Gopakumar, Minwalla and Strominger \cite%
{13}, there are many works to study soliton solutions of
noncommutative field theory and integrable systems in the
background of noncommutative spaces [23-30]. Although Derrick's
theorem forbids solitons in ordinary 2+1 dimensional scalar field
theory \cite{bib:33}, solitons in noncommutative scalar field
theory on the plane were constructed in terms of projection operators in %
\cite{13}. It was soon realized that noncommutative solitons
represent D-branes in string field theory with a background $B$
field, and many of Sen's conjectures \cite{18,19} regarding
tachyon condensation in string field theory have been beautifully
confirmed using properties of noncommutative solitons. Gopakumar,
Minwalla and Strominger makde an important finding that in a
noncommutative space, a projector may correspond to a soliton in
the field theory \cite{13}, which proves the significance of the
study of projection operators in various noncommutative space.
Reiffel \cite{15} constructed the complete set of projection
operators on the noncommutative torus $T^{2}$. On the basis Boca
studied the projection operators on noncommutative orbifold
\cite{16} obtaining many important results and showed the
well-known example of projection operator for $T^{2}/Z_{4}$ in
terms of the elliptic function. Soliton solutions in
noncommutative gauge theory were introduced by Polychronakos in
\cite{soliton}. Martinec and Moore in their important article
deeply studied soliton solutions namely projectors on a wide
variety of orbifolds, and the relation between physics and
mathematics in this area \cite{11}. Gopakumar, Headrick and
Spradlin have shown a rather apparent method to construct the
multi-soliton solution on noncommutative integral torus with generic $\tau$%
\cite{10}. This approach can be generalized to construct the
projection operators on the integral noncommutative orbifold
$T^2/Z_N$ \cite{9}.\newline

In this paper, in the case of integral noncommutative orbifold
$T^2/Z_4$ generated by $u_{1}$ and $u_{2}$ with
\begin{equation}
 u_{1}u_{2}
=u_{2}u_{1}e^{2\pi i/A},~~~A=1,2,3,\cdots
\end{equation}
 we generalize the
$GHS$ construction, presenting the explicit symmetric form of a
series of projectors with manifest $Z_{4}$ symmetry. It includes
all the solutions with minimal trace, and in the standard
expansions for the projectors (see equation (\ref{eq:41}))
\begin{equation}
P=\sum_{s,t}u_1^s u_2^t\Psi_{s,t}(u_1^A,u_2^A)
\end{equation}
where the eigen value function $\Psi_{s,t}(v_1^A,v_2^A)$ is
continuous (where $v^{A}_{1}$ and $v^{A}_{2}$ are eigenvalues of
$u^{A}_{1}$ and $u^{A}_{2}$). The solutions include two arbitrary
complex functions with $Z_4$ symmetry. The kernels of the
integrations are closed analytic functions of $u_{1}$ and $u_{2}$.
In the simplest case, when $A$ is an even number, we reobtain the
Boca's classic result \cite{16} and obtain a new result when $A$
is an odd number. Moreover the above construction is also
applicable to the integral $T^2/Z_N (N=3,6)$ cases.\newline

This paper is organized as following: In Section 2, we introduce
operators
on the noncommutative orbifold $T^2/Z_N$. In Section 3, we introduce the $%
|k,q>$ representation and provide the matrix element relation for
the projectors and deduce the relation between the eigen value
functions of coefficients and the matrix elements of operators in
the $|k,q>$ representation. In Section 4, we study the general
projectors with minimal trace when the eigen value functions of
coefficients are continuous. In Section 5, we present two kinds of
explicit expressions for the projectors with elliptical functions
as kernel.

\section{Noncommutative Orbifold $T^2/Z_N$}

In this section, we introduce operators on the noncommutative orbifold $%
T^{2}/Z_{N}$. First we introduce two hermitian operators $\hat{y_{1}}$ and $%
\hat{y_{2}}$, which satisfy the following commutation relation:
\begin{equation}
\lbrack \hat{y_{1}},\hat{y_{2}}]=i.
\end{equation}%
The operators made up of $\hat{y_{1}}$ and $\hat{y_{2}}$
\begin{equation}
\hat{O}=\sum_{m,n}C_{mn}\hat{y}_{1}^{m}\hat{y}_{2}^{n}
\end{equation}%
form the noncommutative plane $R^{2}$. All operators on $R^{2}$
which commute with $U_{1}$ and $U_{2}$
\begin{equation}
U_{1}=e^{-il\hat{y_{2}}},~~~~~~~~~~U_{2}=e^{il(\tau _{2}\hat{y_{1}}-\tau _{1}%
\hat{y_{2}})},
\end{equation}%
where $l,\tau _{1},\tau _{2}$ are all real numbers and $l,\tau
_{2}>0,\tau =\tau _{1}+i\tau _{2}$, constitute the noncommutative
torus $T^{2}$. We have
\begin{eqnarray}
U_{1}^{-1}\hat{y_{1}}U_{1} &=&\hat{y_{1}}+l,~~~~~~~U_{2}^{-1}\hat{y_{1}}%
U_{2}=\hat{y_{1}}+l\tau _{1},  \nonumber \\
U_{1}^{-1}\hat{y_{2}}U_{1} &=&\hat{y_{2}},~~~~~~~~~~~U_{2}^{-1}\hat{y_{2}}%
U_{2}=\hat{y_{2}}+l\tau _{2}.
\end{eqnarray}%
The operators $U_{1}$ and $U_{2}$ are two different wrapping
operators
around the noncommutative torus and their commutation relation is $%
U_{1}U_{2}=U_{2}U_{1}e^{-2\pi i\frac{l^{2}\tau _{2}}{2\pi }}$. When $A=\frac{%
l^{2}\tau _{2}}{2\pi }$ is an integer, we call the noncommutative
torus integral.Introduce two operators $u_{1}$ and $u_{2}$:
\begin{eqnarray}\label{eq:40}
u_{1} &=&e^{-il\hat{y_{2}}/A},~~~~~~~~~~u_{2}=e^{-il(\tau _{2}\hat{y_{1}}%
-\tau _{1}\hat{y_{2}})/A},  \nonumber  \label{eq:14} \\
u_{1}u_{2} &=&u_{2}u_{1}e^{2\pi
i/A},~~~~~~u_{1}^{A}=U_{1},~~u_{2}^{A}=U_{2}^{-1}
\end{eqnarray}%
The operators on the noncommutative torus are composed of the
Laurant series of $u_{1}$ and $u_{2}$,
\begin{equation}
\hat{O}_{T^{2}}=\sum_{m,n}C_{mn}^{\prime }u_{1}^{m}u_{2}^{n}
\label{eq:5}
\end{equation}%
where $m,n\in Z$ and $C_{00}^{\prime }$is called the trace of the
operators, Eq.(\ref{eq:5}) includes all operators on the
noncommutative torus $T^{2}$, satisfying the relation
$U_{i}^{-1}\hat{O}_{T^{2}}U_{i}=\hat{O}_{T^{2}}$. From
(\ref{eq:40}) we can rewrite the equation(\ref{eq:5}) as
\begin{equation}\label{eq:41}
\hat{O}_{T^{2}}=\sum_{s,t=0}^{A-1}u_{1}^{s}u_{2}^{t}\Psi
_{st}(u_{1}^{A},u_{2}^{A})  \label{eq:6}
\end{equation}%
where $\Psi _{st}$ is Laurant series of the operators $u_{1}^{A}$and $%
u_{2}^{A}$. we call this formula the standard expression for the
operator on the noncommutative torus $T^{2}$. The trace for the
operator is the constant term's coefficient of $\Psi _{00}$.
Nextly we introduce rotation $R$ in noncommutative space $R^{2}$
\begin{equation}
R(\theta )=e^{-i\theta \frac{\hat{y_{1}}^{2}+\hat{y_{2}}^{2}}{2}+i\frac{%
\theta }{2}}
\end{equation}%
with
\begin{equation}
R^{-1}\hat{y}_{1}R =\cos \theta \hat{y}_{1}+\sin \theta
\hat{y}_{2},~~~~ R^{-1}\hat{y}_{2}R =\cos \theta \hat{y}_{2}-\sin
\theta \hat{y}_{1}.
\end{equation}%
When $\tau =\tau _{1}+\tau _{2}=e^{2\pi i/N}$, setting $\theta
=2\pi /N(N\in
Z)$. The noncommutative torus $T^{2}$ keep invariant under rotation $%
R_{N}\equiv R(2\pi /N)$ \cite{16,11,9}. Namely $R_{N}^{-1}\hat{O}%
_{T^{2}}R_{N} $ is still the operators on the noncommutative Torus
$T^{2}$. Now $U_{i}^{\prime }\equiv R_{N}^{-1}U_{i}R_{N}$ can be
expressed by monomial of $\{U_{i}\}$ and their inverses
\cite{11}.In this case, we call the operators invariant under
rotation $R_{N}$ on the noncommutative torus as operators on
noncommutative orbifold $T^{2}/Z_{N}$. We can also realize these
operators in Fock space. Introduce
\begin{equation}
a=\frac{\hat{y}_{2}-i\hat{y}_{1}}{\sqrt{2}},~~~~~~~~~a^{+}=\frac{\hat{y}%
_{2}+i\hat{y}_{1}}{\sqrt{2}},
\end{equation}%
then
\begin{eqnarray}
\lbrack a,a^{+}] &=&1, \\
R_{N} &=&e^{-i\theta a^{+}a}.
\end{eqnarray}%
In this paper, we study the projector $P$ on the orbifold
$T^{2}/Z_{4}$:
\begin{eqnarray}
\tau &=&i, \\
P^{2} &=&P, \\
U_{j}^{-1}PU_{j} &=&P,~~~~~~j=1,2 \\
R_{4}^{-1}PR_{4} &=&P.
\end{eqnarray}

\section{The $|k,q>$ representation, standard form and eigen value function}

From the above discussion, we know that the operators $U_{1}$ and
$U_{2}$ commute with each other on the integral torus $T^{2}$ when
$A$ is an integer. So we can introduce a complete set of their
common eigenstates, namely $|k,q>$ representation \cite{21,20}
\begin{equation}
|k,q>=\sqrt{\frac{l}{2\pi }}e^{-i\tau _{1}\hat{y_{2}}^{2}/2\tau
_{2}}\sum_{j}e^{ijkl}|q+jl>,
\end{equation}%
where the ket on the right is a $\hat{y_{1}}$ eigenstate. We have
\begin{eqnarray}
U_{1}|k,q>= &&e^{-ilk}|k,q>,~~~~~~~U_{2}|k,q>=e^{il\tau
_{2}q}|k,q>=e^{2\pi
iqA/l }|k,q>,  \nonumber  \label{eq:7} \\
id &=&\int_{0}^{\frac{2\pi }{l}}dk\int_{0}^{l}dq|k,q><k,q|.
\end{eqnarray}%
It also satisfies
\begin{equation}\label{eq:46}
|k,q>=|k+\frac{2\pi }{l},q>=e^{ilk}|k,q+l>.
\end{equation}%
Consider the equation (\ref{eq:6}), namely the standard expansion
of operators on $T^{2}$ we have
\begin{equation}\label{eq:15}
\Psi
_{st}(u_{1}^{A},u_{2}^{A})|k,q>=\Psi _{st}(e^{-ilk},e^{-2\pi
iqA/l})|k,q>\equiv \psi _{st}(k,q)|k,q>,
\end{equation}%
where $\psi _{st}$ is a function of the independent variables $k$
and $q$, called
the eigen value function of $\Psi _{st}(u_{1}^{A},u_{2}^{A})$. From (\ref{eq:15}%
), we see that the function $\psi _{st}$ is invariant when
$q\rightarrow q+l/A$,
\begin{equation}
\psi _{st}(k,q+\frac{ln}{A})=\psi _{st}(k,q).  \label{eq:16}
\end{equation}%
As long as the eigen value function is obtained, the operator on
the
noncommutative torus can be completely determined. Introducing new basis $%
|k,q_{0};n>\equiv |k,q_{0}+\frac{ln}{A}>, k\in \lbrack 0,\frac{2\pi }{l}%
), q_{0}\in \lbrack 0,\frac{l}{A})$, we have from (\ref{eq:7})
\begin{equation}
\sum_{n=0}^{A-1}\int_{0}^{\frac{2\pi }{l}}dk\int_{0}^{\frac{l}{A}%
}dq_{0}|k,q_{0}+\frac{ln}{A}><k,q_{0}+\frac{ln}{A}|=id,
\end{equation}%
\begin{equation}
u_{1}|k,q>=|k,q+\frac{l}{A}>,  \label{eq:21}
\end{equation}%
\begin{equation}
u_{2}|k,q>=e^{-2\pi i\frac{q}{l}}|k,q>.  \label{eq:22}
\end{equation}%
From the above equation and (\ref{eq:46}), we see that when any
power of the operators $u_{1}$ and $u_{2}$ act on the
$|k,q_{0}+\frac{ln}{A}>$, the result can be expanded in the basis
$|k,q_{0}+\frac{ln^{\prime }}{A}>$ with the same $k,q_{0}$. So the
operators on the noncommutative torus have the
same property, namely don't change $k$ and $q_{0}$. Thus, for every $k$ and $%
q_{0}$ we get a $A\times A$ matrix, called reduced matrix for the
operator, as well as the projector:
\begin{equation}\label{eq:23}
P_{T^{2}}|k,q_{0}+\frac{ln}{A}>=\sum_{n^{\prime
}}M(k,q_{0})_{n^{\prime }n}|k,q_{0}+\frac{ln^{\prime }}{A}>,
\end{equation}%
It is easy to find that the sufficient and necessary condition for
$P^{2}=P$ is \cite{9}
\begin{equation}
M(k,q_{0})^{2}=M(k,q_{0}).
\end{equation}%
When $T^{2}$ satisfies $Z_{N}$ symmetry, since after $R_{N}$ rotation $%
U_{i}^{\prime }$ can be expressed by monomial of $\{U_{i}\}$ and
their inverses, the state vector $R_{N}|k,q_{0}+\frac{ln}{A}>$ is
still the common
eigenstate of the operators $U_{1}$ and $U_{2}$. With the completeness of $%
\{|k,q+\frac{ls}{A}>\}$ and the A-fold degeneracy eigenvalues of
$U_{i}$ in the $kq$ representation, the state can be expanded in
the basis $\{|k^{\prime },q^{\prime }+\frac{ls^{\prime }}{A}>\}$
\begin{equation}\label{eq:17}
R_{N}|k,q_{0}+\frac{ln}{A}>=\sum_{n^{\prime
}}A(k,q_{0})_{n^{\prime }n}|k,q_{0}+\frac{ln^{\prime }}{A}>
\end{equation}\footnote{It is necessary to point out that the matrix $A$
defined here is the transposed matrix of $A$ defined in Formula (93) in paper \cite{9}. }%
where $k^{\prime }\in \lbrack 0,2\pi /l),q^{\prime }\in \lbrack
0,l/A)$ are definite and
\begin{equation}\label{eq:18}
R_{N}^{-1}|k^{\prime },q_{0}^{\prime }+\frac{ln^{\prime }}{A}%
>=\sum_{n"}A^{-1}(k,q_{0})_{n"n^{\prime }}|k^{\prime },q^{\prime }_{0}+\frac{ln"}{A}>.
\end{equation}%
We can get the expression for the relation between $k^{\prime }$,$%
q_{0}^{\prime }$ and $k$, $q_{0}$, The mapping
$W:(k,q_{0})\longrightarrow (k^{\prime },q_{0}^{\prime
}),W^{N}=id$, is essentially a linear relation, and
area-preserving. By this fact and since $R_{N}$ is unitary, we
conclude that the matrix $A$ is a unitary matrix, that is to say
\begin{equation}
A^{\ast }(k,q_{0})_{nn^{\prime }}=A^{-1}(k,q_{0})_{n^{\prime }n}.
\label{eq:19}
\end{equation}%
The projector on the noncommutative orbifold $T^{2}/Z_{N}$ satisfies $%
R_{N}^{-1}PR_{N}=P$, then from
(\ref{eq:23})(\ref{eq:17})(\ref{eq:18}) one obtains
\begin{equation}\label{eq:35}
R_{N}^{-1}PR_{N}|k,q_{0}+\frac{ln}{A}>=\sum_n^{'}[A^{-1}(k,q_{0})M(k^{\prime
},q_{0}^{\prime })A(k,q_{0})]_{n^{'}n}|k,q_{0}+\frac{ln^{'}}{A}>,
\end{equation}
which should be equal to :
\begin{equation}\label{eq:36}
P|k,q_{0}+\frac{ln}{A}>=\sum_{n"}M(k,q_{0})_{n"n}|k,q_{0}+\frac{ln"}{A}>.
\end{equation}%
So, we have
\begin{equation}
M(k^{\prime },q_{0}^{\prime })=A(k,q_{0})M(k,q_{0})A^{-1}(k,q_{0})
\label{eq:10}
\end{equation}%
and the sufficient and necessary condition for the projector on
noncommutative orbifold $T^{2}/Z_{N}$ to satisfy is:
\begin{eqnarray}
M(k,q_{0})^{2} &=&M(k,q_{0}), \\
M(k^{\prime },q_{0}^{\prime })
&=&A(k,q_{0})M(k,q_{0})A^{-1}(k,q_{0}).\label{eq:24}
\end{eqnarray}%
Next we will study the relation between coefficient function $\psi
_{st}(k,q) $ and the reduced matrix $M(k,q_{0})$. From
(\ref{eq:16})(\ref{eq:21})(\ref{eq:22}) and (\ref{eq:23})we have
\begin{eqnarray}
P|k,q_{0}+\frac{ln}{A}>= &&\sum_{s,t}u_{1}^{s}u_{2}^{t}\Psi
_{st}(u_{1}^{A},u_{2}^{A})|k,q_{0}+\frac{ln}{A}>  \nonumber \\
&=&\sum_{s,t}e^{-2\pi i(q_{0}/l+n/A)t}\psi _{st}(k,q_{0})|k,q_{0}+\frac{%
l(n+s)}{A}>  \nonumber \\
&=&\sum_{n^{\prime }}M(k,q_{0})_{n^{\prime }n}|k,q_{0}+\frac{ln^{\prime }%
}{A}>.
\end{eqnarray}%
So for $n+s<A$ case, we have
\begin{equation}
M(k,q_{0})_{n+s,n}=\sum_{t=0}^{A-1}e^{-2\pi i(q_{0}/l+n/A)t}\psi
_{st}(k,q_{0})
\end{equation}%
and for $n+s\geq A$ case, we have
\begin{equation}
M(k,q_{0})_{n+s-A,n}=\sum_{t=0}^{A-1}e^{-2\pi i(q_{0}/l+n/A)t}\psi
_{st}(k,q_{0})e^{-ilk}.
\end{equation}%
Setting
\begin{equation}
M(k,q_{0})_{n+s,n}=M(k,q_{0})_{n+s-A,n}e^{ilk},  \label{eq:1}
\end{equation}%
We can uniformly write as:
\begin{equation}
M(k,q_{0})_{n+s,n}=\sum_{t=0}^{A-1}e^{-2\pi i(q_{0}/l+n/A)t}\psi
_{st}(k,q_{0})  \label{eq:2}
\end{equation}%
and have
\begin{equation}
\psi
_{st}(k,q_{0})=\frac{1}{A}\sum_{n=0}^{A-1}M(k,q_{0})_{n+s,n}e^{2\pi
i(q_{0}/l+n/A)t}.  \label{eq:3}
\end{equation}%
Eq.(\ref{eq:2}) and (\ref{eq:3}) is the relation between $\psi
_{st}$ and the elements of reduced matrix $M$.

\section{Continuous solution for the Projector with Minimal Trace}

Now one may ask what property the reduced matrix $M$ possess when
the coefficient function $\psi _{st}$ is a continuous function. In
this section, we mainly answer this question. First we prove the
$A\times A$ matrix
satisfying the condition $M^{2}=M$ is always diagonalizable. For any vector $%
\psi $, $M\psi $ is invariant under $M$, namely
\begin{equation}
M(M\psi )=M\psi .
\end{equation}%
Assume there are totally $B$ linear independent invariant vectors
under transformation $M$, then\newline (1) for $A=B$ case, the
matrix $M$ is identity of the space expanded by the vectors,
namely $A\times A$ unit matrix. Of course it is diagonal.\newline
(2) for $B<A$ case, considering any vector $a$ and setting
$b=Ma-a$, we find $Mb=0$. Namely any vector $a$ can be expressed
as linear combination of invariant vector $c=Ma$ and null vector
$b$ under action of $M$. So the whole linear space is composed of
certain invariant vectors and null vectors under action of $M$.
$M$ can be diagonalized in the representation with these vectors
as basis. So we have:
\begin{equation}
M(k,q_{0})=S^{-1}(k,q_{0})\overline{M}(k,q_{0})S(k,q_{0}),
\end{equation}%
where
\begin{equation}
\overline{M}(k,q_{0})=diag(1,1,\cdots ,1,0,0,\cdots ,0).
\end{equation}%
Due to (\ref{eq:2}), when $\psi _{st}(k,q_{0})$ is continuous,
$M(k,q_{0})$ is also continuous. However
$trM(k,q_{0})=tr\overline{M}(k,q_{0})=0,1,2,\cdot \cdot \cdot ,A$,
which is discrete, so when $\psi _{st}$ is continuous, the value
of $trM(k,q_{0})=A\psi_{00}(k,q_{0})$ is invariant for all $k$ and
$q_{0}$. The trace of the projector is the zero order term of
$\psi _{00}(k,q_{0})$ in Laurant expression of $e^{-ilk}$ and
$e^{-2\pi iqA/l}$, so we have
\begin{eqnarray}
trP &=&\int_{0}^{\frac{2\pi }{l}}dk\int_{0}^{\frac{l}{A}}dq\frac{A}{2\pi }%
\psi _{00}(k,q_{0}) \nonumber\\
&=&\int_{0}^{\frac{2\pi }{l}}dk\int_{0}^{\frac{l}{A}}dq\frac{1}{2\pi }%
trM(k,q_{0}) \nonumber\\
&=&\frac{1}{A}trM(k,q_{0}).
\end{eqnarray}%
The projector is trivial for $trM(k,q_{0})=0,A$, indicating $P=0$ and $%
identity$. The nontrivial $trP=\frac{1}{A},\frac{2}{A},\cdots ,\frac{A-1}{A}$%
. In this paper, we only study the nontrivial projector with minimal trace($%
trM(k,q_{0})=1$). Thus

\begin{equation}
M(k,q_0)=s^{-1}(k,q_0) \left(
\begin{array}{cccc}
1 &  &  &  \\
& 0 &  &  \\
&  & 0 &  \\
&  &  & \ddots%
\end{array}
\right) s(k,q_0),  \nonumber
\end{equation}
\begin{equation}  \label{eq:8}
M(k,q_0)_{nn^{\prime}}=s^{-1}(k,q_0)_{n0}
s(k,q_0)_{0n^{\prime}}\equiv a(k,q_0)_{n}b(k,q_0)_{n^{\prime}}.
\end{equation}

Explicit calculation about $R_{N}$ acting on $|k,q;n>$ shows that
we can divide the complete area $\Sigma :\{k\in \lbrack 0,2\pi
/l),q_{0}\in \lbrack 0,l/A)\}$ into $N$ subarea $\sigma
_{0},\cdots ,\sigma _{N-1}$,making $W:\sigma _{i}\rightarrow
\sigma _{i+1},(i=0,1,\cdots ,N-2),\sigma _{N-1}\rightarrow
\sigma _{0}.$ If we construct a reduced matrix $M(k,q_{0})$ to satisfy (\ref%
{eq:8}) in the area $\sigma _{0}$, then the projector
corresponding to continuous $\psi _{st}$ with minimal trace is
completely determined. In area $\sigma _{0}$, set
\begin{equation}
a_{n}=<k,q_{0}+\frac{ln}{A}|\phi _{1}>,~~~~~~~~b_{n}=<\phi _{2}|k,q_{0}+%
\frac{ln}{A}>,  \label{eq:9}
\end{equation}%
where
\begin{equation}
\sum_{n}a_{n}b_{n}=trM(k,q_{0})=1.
\end{equation}%
In the other areas $\sigma_{j}$ with $(k,q_{0})\rightarrow
(k_{j},q_{0j})$ by mapping $W^{j}$, we demand
\begin{eqnarray}
a_{n}(k_{j},q_{0j}) &=&<k_{j},q_{0j}+\frac{ln}{A}|\phi _{1}>  \nonumber \\
&=&A^{j}(k,q_{0})_{nn^{\prime }}a_{n^{\prime }}(k,q_{0}), \\
b_{n}(k_{j},q_{0j}) &=&<\phi _{2}|k_{j},q_{0j}+\frac{ln}{A}>  \nonumber \\
&=&b_{n^{\prime }}(k,q_{0})A^{-j}(k,q_{0})_{n^{\prime }n}.
\end{eqnarray}%
We thus have all coefficients of $|\phi_{1}>, <\phi_{2}|$ in $\sigma_{0},\cdots,\sigma_{N-1}$. Owing to the completeness of $|k,q_{0}+\frac{ln}{A}>$ in the area $%
\Sigma $, $|\phi _{1}>$ and$<\phi _{2}|$ can be determined by the
coefficient (\ref{eq:9}) of $|\phi _{1}>$ and$<\phi _{2}|$.
Meanwhile, in the area $\sigma _{j}$, we have
\begin{eqnarray}
M(k_{j},q_{0j})_{nn^{\prime }} &=&a_{n}(k_{j},q_{0j})b_{n^{\prime}}(k_{j},q_{0j})  \nonumber \\
&=&[A^{j}(k,q_{0})M(k,q_{0})A^{-j}(k,q_{0})]_{nn^{\prime }}.
\end{eqnarray}%
The matrix $M(k,q_{0})$ really satisfies the equation
(\ref{eq:10}). Consider the state vector
\begin{eqnarray}
|\phi _{1}>&= &\int dkdq_{0}\sum_{n}|k,q_{0}+\frac{ln}{A}><k,q_{0}+\frac{ln}{%
A}|\phi _{1}>  \nonumber \\
&=&\sum_{j=0}^{N-1}\int_{\sigma _{j}}dk_{j}dq_{0j}\sum_{n}|k_{j},q_{0j}+\frac{ln}{A}%
>a_{n}(k_{j},q_{0j})  \nonumber \\
&=&\sum_{j=0}^{N-1}\int_{\sigma _{j}}dkdq_{0}\sum_{nn_{1}}|k,q_{0}+\frac{ln}{A}%
>A_{nn_{1}(k,q_{0})}^{j}a_{n_{1}}(k,q_{0})  \nonumber \\
&=&\sum_{j=0}^{N-1}R_{N}^{j}\int_{\sigma _{0}}dkdq_{0}\sum_{n}|k,q_{0}+\frac{ln}{A}%
>a_{n}(k,q_{0}).
\end{eqnarray}%
Thus we have
\begin{equation}  \label{eq:25}
R_{N}|\phi _{1}>=|\phi _{1>}.
\end{equation}%
In the same way, we get
\begin{equation}  \label{eq:26}
<\phi _{2}|R_{N}=<\phi _{2}|.
\end{equation}%
That is to say that the state vectors $|\phi _{1}>$ and $<\phi
_{2}|$ are invariant under the rotation $R_{N}$.\newline

More generally, we can take any state vectors $|\phi _{1}>$ and
$<\phi _{2}|$ satisfying
\begin{equation}\label{eq:50}
R_{N}|\phi _{1}>=e^{i\alpha _{1}}|\phi _{1}>,~~~~~~~<\phi
_{2}|R_{N}^{-1}=e^{-i\alpha _{2}}<\phi _{2}|
\end{equation}%
to construct a projection operator on noncommutative orbifold
$T^{2}/Z_{N}$. Let $M(k,q_{0})$ be given by (\ref{eq:8}) with
\begin{eqnarray}
a_{n}(k,q_{0}) &=&\frac{<k,q_{0}+\frac{ln}{A}|\phi _{1}>}{\sqrt{%
\sum_{n^{\prime }}<k,q_{0}+\frac{ln^{\prime }}{A}|\phi _{1}><\phi
_{2}|k,q_{0}+\frac{ln^{\prime }}{A}>}},  \label{eq:4} \\
b_{n}(k,q_{0}) &=&\frac{<\phi _{2}|k,q_{0}+\frac{ln}{A}>}{\sqrt{%
\sum_{n^{\prime }}<k,q_{0}+\frac{ln^{\prime }}{A}|\phi _{1}><\phi
_{2}|k,q_{0}+\frac{ln^{\prime }}{A}>}}.
\end{eqnarray}%
The projector of minimal trace and with continuous coefficient
functions is surely of this form. It can be verified that
$M^{2}=M$. And it is also covariant under $R_{N}$. From
(\ref{eq:18}) we have

\begin{eqnarray*}
&&<\phi _{2}|k^{^{\prime }},q_{0}^{^{\prime }}+\frac{n^{^{\prime
}}l}{A}>
\\
&=&<\phi _{2}|R_{N}\sum_{n^{"}}A^{-1}(k,q_{0})_{n"n^{\prime
}}|k,q_{0}+\frac{n^{"}l}{A}> \\
&=&e^{i\alpha _{2}}\sum_{n^{"}}A^{-1}(k,q_{0})_{n^{"}n^{^{\prime
}}}<\phi _{2}|k,q_{0}+\frac{n^{"}l}{A}>
\end{eqnarray*}
and similarly

\[
<k^{^{\prime }},q_{0}^{^{\prime }}+\frac{n^{^{\prime }}l}{A}|\phi
_{1}>=e^{-i\alpha _{1}}\sum_{n^{"}}<k,q_{0}+\frac{n^{"}l}{A}|\phi
_{1}>A(k,q_{0})_{n^{^{\prime }}n^{"}},
\]
giving

\begin{eqnarray}
&&\sum_{n}<k^{^{\prime }},q_{0}^{^{\prime }}+\frac{ln}{A}|\phi
_{1}><\phi
_{2}|k^{^{\prime }},q_{0}^{^{\prime }}+\frac{ln}{A}>  \nonumber \\
&=&\sum_{n}<k,q_{0}+\frac{ln}{A}|\phi _{1}><\phi _{2}|k,q_{0}+\frac{ln}{A}%
>e^{-i(\alpha _{1}-\alpha _{2})}.
\end{eqnarray}
Thus
\begin{equation}
M(k^{\prime },q_{0}^{\prime })_{nn^{\prime }}=a_{n}(k^{\prime
},q_{0}^{\prime })b_{n^{\prime }}(k^{\prime },q_{0}^{\prime
})=[AMA^{-1}](k,q_{0})_{nn^{\prime }},
\end{equation}%
$P$ is invariant under rotation $R_{N}$ due to (\ref{eq:24}) and
really gives the projection operator on noncommutative orbifold
$T^{2}/Z_{N}$. The
form of (\ref{eq:4}) is a generalization of $GHS$ construction.\footnote{%
The condition $P^{\dag }=P$ isn't satisfied by $P$ like this,
which might represent the solitons in a ''complex'' field.}. From
the above result, we have
\begin{equation}
M(k,q_{0})_{nn^{\prime }}=\frac{<k,q_{0}+\frac{ln}{A}|\phi
_{1}><\phi
_{2}|k,q_{0}+\frac{ln^{\prime }}{A}>}{\sum_{n^{"}}<k,q_{0}+\frac{ln"}{A}%
|\phi _{1}><\phi _{2}|k,q_{0}+\frac{ln"}{A}>}.
\end{equation}%
Noticing that this equation satisfies (\ref{eq:1}), we have
\begin{eqnarray}  \label{eq:30}
\psi _{st}(k,q_{0})
&=&\frac{1}{A}\sum_{n=0}^{A-1}M(k,q_{0})_{n+s,n}e^{2\pi
i(q_{0}/l+n/A)t}  \nonumber \\
&=&\frac{\frac{1}{A}\sum_{n=0}^{A-1}<k,q_{0}+\frac{l(n+s)}{A}|\phi
_{1}><\phi
_{2}|k,q_{0}+\frac{ln}{A}>e^{2\pi i(q_{0}/l+n/A)t}}{%
\sum_{n}<k,q_{0}+\frac{ln}{A}|\phi _{1}><\phi
_{2}|k,q_{0}+\frac{ln}{A}>}
\nonumber \\
&=&\frac{F_{st}(k,q_{0})}{AF_{00}(k,q_{0})},
\end{eqnarray}%
where
\begin{equation}\label{eq:42}
F_{st}(k,q_{0})\equiv
\sum_{n=0}^{A-1}<k,q_{0}+\frac{l(n+s)}{A}|\phi _{1}><\phi
_{2}|k,q_{0}+\frac{ln}{A}>e^{2\pi i(q_{0}/l+n/A)t},
\end{equation}%
with
\begin{eqnarray}
F_{st}(k,q_{0}) &=&F_{st}(k,q_{0}+l/A)=F_{st}(k+2\pi /l,q_{0}), \\
F_{st}(k,q_{0}) &=&F_{s+A,t}(k,q_{0})e^{-ilk} \\
&=&F_{s,t+A}(k,q_{0})e^{-2\pi iq_{0}A/l}.
\end{eqnarray}%
So the function $F_{st}$ is the function of independent variables $%
X=e^{-ilk} $ and $Y=e^{-2\pi iq_{0}A/l}$, namely
$F_{st}(k,q_{0})=\Phi _{st}(X,Y)$. Similarly
\begin{equation}
\psi _{st}(k,q_{0})=\Psi _{st}(X,Y)=\frac{\Phi _{st}(X,Y)}{A\Phi
_{00}(X,Y)}. \label{eq:12}
\end{equation}%
If we change the variable $X$ and $Y$ into $u_{1}^{A}$ and
$u_{2}^{A}$ respectively, the standard form (\ref{eq:6}) of the
projection operator can be easily obtained. So the key question is
to find out $F_{st}(k,q_{0})$.

\section{ Coherent State Representation}

Introduce coherent states
\[
|z>=e^{-\frac{1}{2}z\bar{z}}e^{a^{+}z}|0>,
\]%
where $z=x+iy,\bar{z}=x-iy$, which satisfies
\begin{equation}
\frac{1}{\pi }\int_{-\infty }^{\infty }d^{2}z|z><z|\equiv \frac{1}{\pi }%
\int_{-\infty }^{\infty }dxdy|z><z|=identity,
\end{equation}
\begin{equation}
R_{N}|z>=|\omega _{N}z>.
\end{equation}

We can show \cite{9}
\begin{equation}  \label{eq:27}
<k,q|z>=\frac{1}{\sqrt{l}\pi ^{1/4}}\theta (\frac{q+\frac{\tau }{\tau _{2}}%
k-i\sqrt{2}z}{l},\frac{\tau }{A})e^{-\frac{\tau }{2i\tau _{2}}k^{2}+ikq+%
\sqrt{2}kz-(z^{2}+z\bar{z})/2},
\end{equation}%
where

\bigskip
\[
\theta (z,\tau )\equiv \theta \left[
\begin{array}{c}
0 \\
0%
\end{array}%
\right] (z,\tau )
\]

and%
\begin{equation}  \label{eq:34}
\theta \left[
\begin{array}{c}
a \\
b%
\end{array}%
\right] (z,\tau )=\sum_{m}e^{\pi i\tau (m+a)^{2}}e^{2\pi
i(m+a)(z+b)}.
\end{equation}%
Thus we can expand the state vectors $|\phi _{1}>$ and $<\phi
_{2}|$ in terms of coherent state,
\begin{eqnarray}
|\phi _{1}>&= &\frac{1}{\pi }\int_{-\infty }^{\infty
}dxdy|z><z|\phi _{1}>
\nonumber  \label{eq:28} \\
&\equiv &\frac{1}{\pi }\int_{-\infty }^{\infty }dxdyf_{1}(z)|z>, \\
<\phi _{2}| &=&\frac{1}{\pi }\int_{-\infty }^{\infty }dxdy<\phi
_{2}|z><z|
\nonumber \\
&=&\frac{1}{\pi }\int_{-\infty }^{\infty }dxdyf_{2}(z)<z|.
\label{eq:29}
\end{eqnarray}

\bigskip

The condition (\ref{eq:50}) is satisfied if and only if
\begin{equation}
f_{1}(\omega_{N}^{-1}z)=f_{1}(z)e^{i\alpha _{1}},
\end{equation}%
\begin{equation}
f_{2}(\omega_{N}^{-1}z)=f_{2}(z)e^{-i\alpha _{2}}.
\end{equation}

Here $\omega_{N}=e^{-i\frac{2\pi }{N}}$. We have
\begin{eqnarray}  \label{eq:31}
F_{st}(k,q_{0}) &=&\frac{1}{\pi ^{2}}\sum_{n=0}^{A-1}\int <k,q_{0}+\frac{%
l(n+s)}{A}|z_{1}>f_{1}(z_{1})dx_{1}dy_{1}  \nonumber \\
&&\times \int
<z_{2}|k,q_{0}+\frac{ln}{A}>f_{2}(z_{2})dx_{2}dy_{2}\times e^{2\pi
i(q_{0}/l+n/A)t}  \nonumber \\
&=&\frac{1}{\pi ^{2}}\int
dx_{1}dy_{1}dx_{2}dy_{2}g_{st}(k,q_{0},z_{1},z_{2})f_{1}(z_{1})f_{2}(z_{2}),
\end{eqnarray}%
where
\begin{equation}
g_{st}(k,q_{0},z_{1},z_{2})=\sum_{n=0}^{A-1}<k,q_{0}+\frac{l(n+s)}{A}%
|z_{1}><z_{2}|k,q_{0}+\frac{ln}{A}>e^{2\pi i(q_{0}/l+n/A)t}.
\end{equation}%
We call the kernel $g$ as generating function in coherent state
representation. Next, we study the expression of $g$ for $Z_{4}$
case. Through $g$ we can give the integration expression for all
the projection operators on the $T^{2}/Z_{4}$ with minimal trace
and continuous eigen value function. Consider the equation
\begin{equation}
\theta (z,\tau )^{\ast }=\theta (z^{\ast },-\tau ^{\ast }).
\end{equation}%
For the $z_{4}$ case, $\tau =i,A=\frac{l^{2}}{2\pi }$, from (\ref{eq:27})(%
\ref{eq:28}) and (\ref{eq:29}) we get
\begin{eqnarray}
&<&k,q+\frac{ls}{A}|z_{1}><z_{2}|k,q+\frac{ls^{\prime }}{A}>  \nonumber \\
&=&\frac{1}{l\sqrt{\pi }}\theta (\frac{q}{l}+\frac{i}{l}k+\frac{s}{A}-\frac{i%
\sqrt{2}z_{1}}{l},\frac{i}{A})\theta (\frac{q}{l}-\frac{i}{l}k+\frac{%
s^{\prime }}{A}+\frac{i\sqrt{2}z_{2}^{\ast }}{l},\frac{i}{A})  \nonumber \\
&&\times e^{-k^{2}+\frac{k}{2}(z_{1}+z_{2}^{\ast })+ik\frac{l(s-s^{\prime })%
}{A}}e^{-\frac{1}{2}(z_{1}^{2}+(z_{2}^{\ast
})^{2}+z_{1}z_{1}^{\ast }+z_{2}z_{2}^{\ast })}\equiv K_{ss^{\prime
}}.
\end{eqnarray}%
Let $u=\frac{lk}{2\pi },v=\frac{q}{l},\mu =-i\frac{\sqrt{2}A}{l}z_{1},\nu =i%
\frac{\sqrt{2}A}{l}z_{2}^{\ast }$, then
\begin{eqnarray}
K_{ss^{\prime }} &=&\frac{C_{1}}{l\sqrt{\pi }}\theta (v+\frac{iu}{A}+\frac{%
s+\mu }{A},\frac{i}{A})\theta (v-\frac{iu}{A}+\frac{s^{\prime }+\nu }{A},%
\frac{i}{A})  \nonumber \\
&&\times e^{\pi i\frac{2i}{A}u^{2}+2\pi i(\frac{s+\mu -s^{\prime }-\nu }{A}%
)u}.
\end{eqnarray}%
where
\begin{equation}
C_{1}=e^{2\pi i[\frac{-i}{4A}(\mu ^{2}+\nu ^{2})+\frac{i}{4A}(|\mu
|^{2}+|\nu |^{2})]}.
\end{equation}%
It can be proved that for integer $A$:
\begin{eqnarray}
&&\sum_{r=0}^{A-1}e^{2\pi irt/A}\theta (x+r/A,\tau /A)\theta
(y+r/A,\tau /A)
\nonumber \\
&=&A\sum_{d=0,1}\theta (-\frac{\tau }{A}(Ad-t)+x-y,\frac{2\tau }{A})\theta (%
\frac{\tau }{A}(-At+A^{2}d)+A(x+y),2\tau A)  \nonumber \\
&&\times e^{\pi i\frac{\tau }{A}(Ad-t)^{2}}\times e^{2\pi
i(Ad-t)y}
\end{eqnarray}%
and
\begin{equation}
\theta (z,\tau )=\sqrt{\frac{i}{\tau }}e^{-\pi iz^{2}/\tau }\theta
(\pm \frac{z}{\tau },-\frac{1}{\tau }).
\end{equation}%
Thus we have
\begin{eqnarray}  \label{eq:32}
G_{st}(u,v) &\equiv &g_{st}(k,q,z_{1},z_{2})  \nonumber \\
&=&e^{2\pi ivt}\sum_{r}e^{2\pi irt/A}K_{s+r,r}(u,v)  \nonumber \\
&=&\frac{AC_{1}}{l\sqrt{\pi }}\sqrt{\frac{A}{2}}\sum_{d=0,1}e^{-\frac{\pi }{%
2A}(Ad-t)^{2}+2\pi i(Ad-t)\frac{\nu }{A}-\frac{\pi }{2A}(s+\mu
-\nu )^{2}}
\nonumber \\
&&\times e^{\frac{\pi i}{A}(s+\mu -\nu )(Ad-t)+2\pi ivAd}\theta (u-\frac{1}{2%
}(Ad-t)+\frac{i}{2}(s+\mu -\nu ),\frac{Ai}{2})  \nonumber \\
&&\times \theta (2Av+s+\mu +\nu -t\tau +iAd,2Ai).
\end{eqnarray}%
Due to
\begin{eqnarray}
\sum_{a=0,1}\theta (x+\frac{a}{2},\tau ) &=&2\theta (2x,4\tau ), \\
\sum_{a=0,1}(-1)^{a}\theta (x+\frac{a}{2},\tau ) &=&2e^{2\pi
i(x+\frac{\tau }{2})}\theta (2x+2\tau ,4\tau ),
\end{eqnarray}%
when $d$ is equal to $0,1$,
\begin{equation}
\theta (2x+2\tau d,4\tau )=\frac{1}{2}\sum_{a=0,1}(-1)^{ad}e^{2\pi i(x+\frac{%
\tau }{2})d}\theta (x+\frac{a}{2},\tau ).
\end{equation}%
The function $G_{st}(u,v)$ can be rewritten as
\begin{eqnarray}
G_{st}(u,v) &=&\frac{A}{4\pi }e^{2\pi i\phi
}\sum_{a,d=0,1}(-1)^{ad}\theta
(-u+\frac{1}{2}(Ad+t)+\frac{i}{2}(s+\mu -\nu ),\frac{Ai}{2})
\nonumber
\label{eq:13} \\
&&\times \theta (-Av+\frac{1}{2}(s+a-\mu -\nu
)+\frac{i}{2}t,\frac{Ai}{2}),
\end{eqnarray}%
where
\begin{equation}
\phi =\frac{i}{4A}(s^{2}+t^{2})-\frac{st}{2A}+\frac{i}{2A}s(\mu -\nu )-\frac{%
t}{2A}(\mu +\nu )+\frac{i}{4A}(|\mu |^{2}+|\nu |^{2}-2\mu \nu )
\end{equation}%
From the above discussion, we know that when $A$ is an even number, only $%
a=0 $ contributes, so we have
\[
G_{st}(u,v)=\frac{A}{2\pi }e^{2\pi i\phi}\theta (-u+\frac{1}{2}t+\frac{i%
}{2}(s+\mu -\nu ),\frac{Ai}{2})\theta (-Av+\frac{1}{2}(s-\mu -\nu )+\frac{i}{2}t,\frac{Ai%
}{2})
\]%
and when $A$ is an odd number,
\begin{eqnarray}
G_{st}(u,v) &=&\frac{A}{4\pi }e^{2\pi i\phi
}\sum_{a,d=0,1}(-1)^{ad}\theta
(-u+\frac{1}{2}(Ad+t)+\frac{i}{2}(s+\mu -\nu ),\frac{Ai}{2})  \nonumber \\
&&\times \theta (-Av+\frac{1}{2}(Aa+s-\mu -\nu
)+\frac{i}{2}t,\frac{Ai}{2}).
\end{eqnarray}%
They can be uniformly written as

\begin{eqnarray}
G_{st}(u,v) &=&\frac{A}{4\pi }e^{2\pi i\phi
}\sum_{a,d=0,1}(-1)^{ad}\theta
(-u+\frac{1}{2}(Ad+t)+\frac{i}{2}(s+\mu -\nu ),\frac{Ai}{2})  \nonumber \\
&&\times \theta (-Av+\frac{1}{2}(Aa+s-\mu -\nu
)+\frac{i}{2}t,\frac{Ai}{2}),
\end{eqnarray}%
which is \bigskip $Z_{4}$ covariant. So we have from
(\ref{eq:30})(\ref{eq:31}) and (\ref{eq:32}),
\begin{equation}
\psi _{st}(k,q)=\frac{1}{A}\{\frac{\int d^{2}\mu d^{2}\nu
G_{st}(u,v)f_{1}(\mu )f_{2}(\nu )}{\int d^{2}\mu d^{2}\nu
G_{00}(u,v)f_{1}(\mu )f_{2}(\nu )}\} \label{eq:33}
\end{equation}%
where functions $f_{i}$ should satisfy%
\begin{equation}
f_{i}(\omega _{N}\xi )=e^{i\alpha _{i}}f_{i}(\xi ).
\end{equation}
Let $\hat{u}=\frac{l}{2\pi }\hat{y}_{2}$ and $A\hat{v}=\frac{l}{2\pi }\hat{y}%
_{1}$, we may replace $u,v$ by $\hat{u}$ and $\hat{v}$ in
(\ref{eq:33}) and get
\begin{equation}
\Psi _{st}(u_{1}^{A},u_{2}^{A})=\frac{1}{A}\frac{\int d^{2}\mu
d^{2}\nu G_{st}(\frac{l}{2\pi }\hat{y}_{2},\frac{l}{2A\pi
}\hat{y}_{1})f_{1}(\mu
)f_{2}(\nu )}{\int d^{2}\mu d^{2}\nu G_{00}(\frac{l}{2\pi }\hat{y}_{2},\frac{%
l}{2A\pi }\hat{y}_{1})f_{1}(\mu )f_{2}(\nu )}.
\end{equation}%
The operators $u_{1}$ and $u_{2}$ commute with the operators
$u_{1}^{A}$ and $u_{2}^{A}$,and from (\ref{eq:40})
\begin{eqnarray}
u_{1}^{s} &=&e^{-2\pi i\frac{s}{A}\hat{u}}, \\
u_{2}^{t} &=&e^{-2\pi it\hat{v}}.
\end{eqnarray}%
Further takeing $u_{1}^{s}$ and $u_{2}^{t}$ into account, we can
insert them to the corresponding operator form of
Eq.(\ref{eq:13}). This leads to the function of $\hat{u}$ and
$\hat{v}$
\begin{eqnarray}
h_{ad} &\equiv &u_{1}^{s}u_{2}^{t}G_{st}(\hat{u},\hat{v})\nonumber\\
&=&\frac{A}{4\pi}\sum_{a,d=0}^{1}(-1)^{ad}e^{2\pi i\phi}e^{-2\pi i\frac{s}{A}\hat{u}}\theta (-%
\hat{u}+\frac{1}{2}(Ad+t)+\frac{i}{2}(s+\mu -\nu ),\frac{Ai}{2})
\nonumber
\\
&&\times e^{-2\pi i\hat{v}t}\theta (-A\hat{v}+\frac{1}{2}(Aa+s-\mu -\nu )+\frac{i}{%
2}t,\frac{Ai}{2})  \nonumber \\
&=&\frac{A}{4\pi}\sum_{a,d=0}^{1}(-1)^{ad}e^{2\pi
i(-\frac{3st}{2A}+\frac{i}{4A}(|\mu |^{2}+|\nu |^{2}-2\mu \nu
))}e^{-2\pi i(\frac{sd}{2}+\frac{ta}{2})}  \nonumber \\
&&\times \theta \left[
\begin{array}{l}
\frac{s}{A} \\
\frac{t}{2}%
\end{array}%
\right] (-\hat{u}+\frac{1}{2}Ad+\frac{i}{2}(\mu -\nu
),\frac{Ai}{2})
\nonumber \\
&&\times \theta \left[
\begin{array}{l}
\frac{t}{A} \\
\frac{s}{2}%
\end{array}%
\right] (-A\hat{v}+\frac{1}{2}Aa-\frac{1}{2}(\mu +\nu
),\frac{Ai}{2}).
\end{eqnarray}%
Due to $|\omega _{N}\mu |=|\mu |,|\omega _{N}\nu |=|\nu |$,
$e^{2\pi i\frac{i}{4A}(|\mu |^{2}+|\nu |^{2})}$ in the above
formula can be attributed to $f_{1}(\mu )$ and $f_{2}(\nu )$.
Finally, we have
\begin{eqnarray}
P &=&\sum_{s,t}u_{1}^{s}u_{2}^{t}\Psi _{st}(u_{1}^{A},u_{2}^{A})
\nonumber
\\
&=&\sum_{s,t}e^{2\pi
i(-\frac{3st}{2A})}\sum_{a,d=0}^{1}(-1)^{ad}\int d^{2}\mu
d^{2}\nu f_{1}(\mu )f_{2}(\nu )e^{\pi i\frac{\mu \nu }{A}}e^{-2\pi i(\frac{sd%
}{2}+\frac{ta}{2})}  \nonumber \\
&&\times \theta \left[
\begin{array}{l}
\frac{s}{A} \\
\frac{t}{2}%
\end{array}%
\right] (-\frac{l\hat{y}_{2}}{2\pi }+\frac{1}{2}Ad+\frac{i}{2}(\mu -\nu ),%
\frac{Ai}{2})\theta \left[
\begin{array}{l}
\frac{t}{A} \\
\frac{s}{2}%
\end{array}%
\right] (-\frac{l\hat{y}_{1}}{2\pi }+\frac{1}{2}Aa-\frac{1}{2}(\mu +\nu ),%
\frac{Ai}{2})  \nonumber \\
&&\times \{A\sum_{a,d=0}^{1}(-1)^{ad}\int d^{2}\mu d^{2}\nu
f_{1}(\mu )f_{2}(\nu
)e^{\pi i\frac{\mu \nu }{A}}\theta (-\frac{l\hat{y}_{2}}{2\pi }+\frac{1}{2}%
Ad+\frac{i}{2}(\mu -\nu ),\frac{Ai}{2})  \nonumber \\
&&\times \theta (-\frac{l\hat{y}_{1}}{2\pi
}+\frac{1}{2}Aa-\frac{1}{2}(\mu +\nu ),\frac{Ai}{2})\}^{-1}.
\end{eqnarray}%
In the above equation, the two $\theta $ functions can not
exchange orders with each other. It holds for any integer number
$A$. In the following, we present some discussion.\newline (1) $A$
is an even number, so $\frac{Ad}{2}$ and $\frac{Aa}{2}$ are
integers too. Due to (\ref{eq:34}) we have\newline
\begin{eqnarray}
P &=&\sum_{s,t}e^{-3\pi i\frac{st}{A}}\int d^{2}\mu d^{2}\nu
f_{1}(\mu
)f_{2}(\nu )e^{\pi i\frac{\mu \nu }{A}}  \nonumber \\
&&\times \theta \left[
\begin{array}{l}
\frac{s}{A} \\
\frac{t}{2}%
\end{array}%
\right] (-\frac{l\hat{y}_{2}}{2\pi }+\frac{i}{2}(\mu -\nu ),\frac{Ai}{2}%
)\theta \left[
\begin{array}{l}
\frac{t}{A} \\
\frac{s}{2}%
\end{array}%
\right] (-\frac{l\hat{y}_{1}}{2\pi }-\frac{1}{2}(\mu +\nu
),\frac{Ai}{2})
\nonumber \\
&&\times \{A\int d^{2}\mu d^{2}\nu f_{1}(\mu )f_{2}(\nu )e^{\pi
i\frac{\mu
\nu }{A}}\theta (-\frac{l\hat{y}_{2}}{2\pi }+\frac{i}{2}(\mu -\nu ),\frac{Ai%
}{2})  \nonumber \\
&&\times \theta (-\frac{l\hat{y}_{1}}{2\pi }-\frac{1}{2}(\mu +\nu ),\frac{Ai%
}{2})\}^{-1}.
\end{eqnarray}%
the above equation is the generalization of the Boca's formula Proposition $%
3.1(i)$ \cite{16}.\newline (2) $A$ is an odd number \newline
\begin{eqnarray}
P &=&\sum_{s,t=0}^{A-1}e^{-3\pi i\frac{st}{A}}\int d^{2}\mu
d^{2}\nu
f_{1}(\mu )f_{2}(\nu )e^{\pi i\frac{\mu \nu }{A}}  \nonumber \\
&&\times \sum_{a,d=0}^{1}(-1)^{ad}e^{-2\pi i(\frac{sd}{2}+\frac{ta}{2})}\theta %
\left[
\begin{array}{l}
\frac{s}{A} \\
\frac{t}{2}%
\end{array}%
\right] (-\frac{l\hat{y}_{2}}{2\pi }+\frac{Ad}{2}+\frac{i}{2}(\mu -\nu ),%
\frac{Ai}{2})\nonumber\\
&&\times\theta \left[
\begin{array}{l}
\frac{t}{A} \\
\frac{s}{2}%
\end{array}%
\right] (-\frac{l\hat{y}_{1}}{2\pi }+\frac{Aa}{2}-\frac{1}{2}(\mu +\nu ),%
\frac{Ai}{2})  \nonumber \\
&&\times \{A\int d^{2}\mu d^{2}\nu f_{1}(\mu )f_{2}(\nu )e^{\pi
i\frac{\mu
\nu }{A}}\sum_{a,d=0}^{1}(-1)^{ad}\theta (-\frac{l\hat{y}_{2}}{2\pi }+\frac{Ad}{2}+%
\frac{i}{2}(\mu -\nu ),\frac{Ai}{2})  \nonumber \\
&&\times \theta (-\frac{l\hat{y}_{1}}{2\pi
}+\frac{Aa}{2}-\frac{1}{2}(\mu +\nu ),\frac{Ai}{2})\}^{-1}.
\end{eqnarray}%
Due to
\begin{eqnarray}
&&\theta \left[
\begin{array}{l}
\frac{s}{A} \\
\frac{t}{2}%
\end{array}%
\right] (x+\frac{Ad}{2},\tau )  \nonumber \\
&=&\theta \left[
\begin{array}{l}
\frac{s}{2A} \\
0%
\end{array}%
\right] (2x,4\tau )(-1)^{sd}e^{2\pi i\frac{st}{2A}}+\theta \left[
\begin{array}{l}
\frac{s+A}{2A} \\
0%
\end{array}%
\right] (2x,4\tau )(-1)^{(s+A)d}e^{2\pi i\frac{(s+A)t}{2A}}
\end{eqnarray}%
the numerator of $P$ can be written as
\begin{eqnarray}
&&2\int d^{2}\mu d^{2}\nu f_{1}(\mu )f_{2}(\nu )e^{\pi i\frac{\mu \nu }{A}%
}\sum_{s,t=0}^{A-1}\{\bar{\theta}_{0}\theta
_{0}++\bar{\theta}_{1}\theta _{0}(-1)^{s}+\bar{\theta}_{0}\theta
_{1}(-1)^{t}+\bar{\theta}_{1}\theta
_{1}(-1)^{s+t-1}\}  \nonumber \\
&=&2\int d^{2}\mu d^{2}\nu f_{1}(\mu )f_{2}(\nu )e^{\pi i\frac{\mu
\nu }{A} }\sum_{s,t=0}^{2A-1}e^{\frac{\pi
i}{A}st}\bar{\theta}_{0}\theta _{0}
\end{eqnarray}%
Where $\bar{\theta}_{\delta }=\theta \left[
\begin{array}{c}
\frac{t+A\delta }{2A} \\
0%
\end{array}%
\right] (2y,4\tau ),\theta _{\delta }=\theta \left[
\begin{array}{c}
\frac{t+A\delta }{2A} \\
0%
\end{array}%
\right] (2x,4\tau ),\delta =0,1$ with $x=-\frac{l\hat{y}_{2}}{2\pi }+\frac{i%
}{2}(\mu -\nu )$ and $y=-\frac{l\hat{y}_{1}}{2\pi
}-\frac{1}{2}(\mu +\nu )$. The denominator of $P$ is $2A\int
d^{2}\mu d^{2}\nu f_{1}(\mu )f_{2}(\nu )e^{\pi i\frac{\mu \nu }{A}
}\sum_{\delta_1,\delta_2 =0}^{1}e^{\pi
i\delta_1\delta_2}\bar{\theta}_{\delta_1}^{'}\theta
_{\delta_2}^{'}$, here $\theta^{\prime} _{\delta }=\theta \left[
\begin{array}{c}
\frac{\delta }{2} \\
0%
\end{array}%
\right] (2x,4\tau )$, $\bar{\theta}^{\prime} _{\delta }=\theta
\left[
\begin{array}{c}
\frac{\delta }{2} \\
0%
\end{array}%
\right] (2y,4\tau )$. This formula gives another result compared
with the Boca's when $f_{1}(z)=f_{2}(z)=\delta ^{2}(z-0).$
\bigskip

Finally, we will give another explicit form of $P$ in terms of the
derivative of elliptic functions. Note that the basis $\{|n>\}$ of
Fock space produce a phase $\omega ^{n}$ under action of $R_{N}$.
It is not
difficult to find that $e^{i\alpha _{1}}$ and $e^{i\alpha _{2}}$ in (\ref%
{eq:50}) are both integral powers of $\omega _{N}$ because of $%
(R_{N})^{N}=identity.$ Therefore $|\phi _{1}>$ and $<\phi _{2}|$
can
respectively be expanded in the basis $\{|n>\}$ and $\{<n|\}$, where%
\begin{equation}
|n>=\frac{(a^{+})^{n}}{\sqrt{n!}}|0>.
\end{equation}%
We have the relation between coherent state and particle number
eigenstate
as following:%
\begin{equation}
<z|n>=e^{-\frac{1}{2}z\bar{z}}\frac{z^{n}}{\sqrt{n!}}.
\label{eq:43}
\end{equation}

Obviously, the general forms of $|\phi _{1}>$ and $<\phi _{2}|$ in
the expansion in terms of particle number eigenstates are
$\sum_{m=0}^{\infty }c_{m}$ $|i+4m>$ and $\sum_{n=0}^{\infty
}d_{n}$ $<j+4n|,$ where $i$ and $j$ are nonnegative integers and
$c_{m}$ and $d_{n}$ are arbitrary constant
coefficients. So%
\begin{eqnarray}
|\phi _{1} >&=&\sum_{m}c_{m}|i+4n>  \nonumber \\
&=&\frac{1}{\pi }\sum_{m}\int_{-\infty }^{\infty
}dxdyc_{m}|z><z|i+4m>, \label{eq:44}
\end{eqnarray}%
\begin{eqnarray}
<\phi _{2}|&=&\sum_{m=0}^{\infty }d_{m}<j+4m|  \nonumber \\
&=&\frac{1}{\pi }\sum_{n=0}^{\infty }\int_{-\infty }^{\infty
}dxdyd_{n}<j+4n|z><z|.  \label{eq:45}
\end{eqnarray}%
We let $R_{N}$ act on $|\phi _{1}>$ and $<\phi _{2}|$ and get
\begin{equation}
R_{N}|\phi _{1}>=\omega ^{i}|\phi _{1}>,
\end{equation}

\begin{equation}
<\phi _{2}|R_{N}^{-1}=<\phi _{2}|\omega ^{-j}.
\end{equation}%
Subsequently, we substitute (\ref{eq:43})(\ref{eq:44})(\ref{eq:45}) into (\ref{eq:42})%
 and make use of the formula

\begin{equation}
<k,q|n>=\frac{1}{\sqrt{n!}}\frac{d^{n}}{dz^{n}}\left( e^{\frac{1}{2}z\bar{z}%
}<k,q|z>\right) \Vert _{z=0}
\end{equation}%
to obtain%
\begin{eqnarray}
F_{st}(k,q_{0}) &=&\sum_{m,n}\sum_{h=0}^{A-1}c_{m}d_{n}<k,q_{0}+\frac{l(h+s)%
}{A}|i+4m>\times <j+4n|k,q_{0}+\frac{lh}{A}>\times e^{2\pi
i(q_{0}/l+h/A)t}
\nonumber\\
&=&\sum_{m,n}\sum_{h=0}^{A-1}c_{m}d_{n}\frac{1}{\sqrt{(i+4m)!(j+4n)!}}\frac{%
d^{n+m}}{dz_{1}^{m}\bar{z_{2}}^{n}}(e^{\frac{1}{2}(z_{1}\bar{z}_{1}+z_{2}\bar{z}%
_{2})}\sum_{h=0}^{A-1}<k,q_{0}+\frac{l(h+s)}{A}|z_{1}> \nonumber\\
  &&\times<z_{2}|k,q_{0}+\frac{lh}{A}>\times e^{2\pi
i(q_{0}/l+h/A)t})\Vert
_{z_{1}=\bar{z}_{2}=0} \nonumber\\
&=&\sum_{m,n}\sum_{h=0}^{A-1}c_{m}d_{n}\frac{1}{\sqrt{(i+4m)!(j+4n)!}}\frac{%
d^{n+m}}{dz_{1}^{m}\bar{z_{2}}^{n}}\left( e^{\frac{1}{2}(z_{1}\bar{z}_{1}+z_{2}%
\bar{z}_{2})}g_{st}(k,q_{0},z_{1},z_{2})\right) \Vert _{z_{1}=\bar{z}%
_{2}=0}.\nonumber\\
\end{eqnarray}%
So, we get the projector in the case of $z_{4}$
\begin{eqnarray}
P &=&\sum_{m,n}c_{m}d_{n}\frac{1}{\sqrt{(i+4m)!(j+4n)!}}\frac{d^{n+m}}{dz_{1}^{m}%
\bar{z}_{2}^{n}}(e^{z_{1}\bar{z}_{1}+z_{2}\bar{z}_{2}}\times e^{2\pi i(-%
\frac{3st}{2A})}\times e^{4\pi ^{2}iz_{1}\bar{z}_{2}}\times e^{-2\pi i(%
\frac{sd}{2}+\frac{ta}{2})} \nonumber\\
&&\times \sum_{a,d=0}^{1}(-1)^{ad}\theta \left[
\begin{array}{l}
\frac{s}{A} \nonumber\\
\frac{t}{2}%
\end{array}%
\right] (-\frac{l\hat{y}_{2}}{2\pi }+\frac{1}{2}Ad+\frac{\sqrt{2}A}{2l}%
(z_{1}+\bar{z}_{2}),\frac{Ai}{2}) \nonumber\\
&&\times \theta \left[
\begin{array}{l}
\frac{t}{A} \nonumber\\
\frac{s}{2}%
\end{array}%
\right] (-\frac{l\hat{y}_{1}}{2\pi }+\frac{1}{2}Aa+\frac{i\sqrt{2}A}{2l}%
(z_{1}-\bar{z}_{2}),\frac{Ai}{2}))\Vert _{z_{1}=\bar{z}_{2}=0} \nonumber\\
&&\times \{A\sum_{m,n}\sum_{a,d=0}^{1}(-1)^{ad}c_{m}d_{n}\frac{d^{n+m}}{%
dz_{1}^{m}\bar{z}_{2}^{n}}(e^{z_{1}\bar{z}_{1}+z_{2}\bar{z}_{2}}\times
e^{4\pi
^{2}iz_{1}\bar{z}_{2}}\times \theta (-\frac{l\hat{y}_{2}}{2\pi }+\frac{1%
}{2}Ad+\frac{\sqrt{2}A}{2l}(z_{1}+\bar{z}_{2}),\frac{Ai}{2}) \nonumber\\
&&\times \theta (-\frac{l\hat{y}_{1}}{2\pi }+\frac{1}{2}Aa+\frac{i\sqrt{2}A}{%
2l}(z_{1}-\bar{z}_{2}),\frac{Ai}{2}))\}^{-1}\Vert _{z_{1}=\bar{z}%
_{2}=0}.
\end{eqnarray}

Thus, We derive two forms of explicit expressions of the projector
$P$ in terms of the integration and derivative of the classical
theta functions.

\section{Discussion}

In this paper, $P$ is represented by a form of fraction which make
sense only when the denominator has inverse. The formula demands:
\begin{equation}
D=A\int d^{2}\mu d^{2}\nu f_{1}(\mu )f_{2}(\nu )G_{00}(u,v)
\end{equation}%
is unequal to zero for any real variables $u$ and $v$. It is easy
to prove that when $f_{1}$ is equal to $f_{2}^{\ast }$, the
related denominator
\begin{equation}
D=A\sum_{n}<k,q+\frac{ln}{A}|\phi ><\phi |k,q+\frac{ln}{A}>=A\sum_{n}|<k,q+%
\frac{ln}{A}|\phi >|^{2}.
\end{equation}%
Thus if $D=0$,then
\begin{equation}
<k,q+\frac{ln}{A}|\phi >=0~~~~~~~~n=0,1,\cdot \cdot \cdot ,A-1.
\label{eq:11}
\end{equation}%
The zero points of the state vector $|\phi >$ in $|k,q>$
representation should be points equally spaced along $q$ with interval of $%
\frac{l}{A}$. The mapping from $k$ and $q$ to
$<k,q+\frac{ln}{A}|\phi >\in C$ is a mapping from plane to plane.
In general, $<k,q+\frac{ln}{A}|\phi >=0$ are some discrete points,
and thus it is casual that $D$ is equal to zero. So in this sense,
for most of $f_{1}=f_{2}^{\ast }$, this still not happen (in some
sense, the measure of $D=0$ event is zero.) Specially, when the
state $|\phi _{1}>=|\phi _{2}>=|0>$, It can be proved \cite{16}
that $D$ is not equal to zero everywhere. Thus set
\begin{equation}
|\phi _{1}>=|0>+\epsilon |\psi _{1}>,~~~~~~~<\phi
_{2}|=<0|+\epsilon <\psi _{2}|
\end{equation}%
$D$ is also not equal to zero everywhere for small enough
$\epsilon $. But we don't know the situation for general
$f_{1}\neq f_{2}^{\ast }.$

\end{document}